\documentclass[twocolumn,runningheads,natbib]{svjour2}
\smartqed  
\usepackage{graphicx}
\journalname{Astrophysics and Space Science}
\begin{document}

\title{Internal heating and thermal emission from old neutron stars} \subtitle{Constraints on dense-matter and gravitational physics}


\author{Andreas Reisenegger \and
        Rodrigo Fern\'andez \and
        Paula Jofr\'e
}


\institute{A. Reisenegger and P. Jofr\'e \at
              Departamento de Astronom\'\i a y Astrof\'\i sica \\
              Pontificia Universidad Cat\'olica de Chile \\
              Casilla 306, Santiago 22, Chile \\
              \email{areisene@astro.puc.cl;pdjofre@uc.cl}
           \and
           R. Fern\'andez \at
              Department of Astronomy \& Astrophysics \\
              University of Toronto \\
              Toronto, ON M5S 3H8, Canada \\
              \email{fernandez@astro.utoronto.ca}
}

\date{Received: date / Accepted: date}

\maketitle

\begin{abstract}
The equilibrium composition of neutron star matter is achieved
through weak interactions (direct and inverse beta decays), which
proceed on relatively long time scales. If the density of a matter
element is perturbed, it will relax to the new chemical equilibrium
through non-equilibrium reactions, which produce entropy that is
partly released through neutrino emission, while a similar fraction
heats the matter and is eventually radiated as thermal photons. We
examined two possible mechanisms causing such density perturbations:
1) the reduction in centrifugal force caused by spin-down
(particularly in millisecond pulsars), leading to {\it rotochemical
heating}, and 2) a hypothetical time-variation of the gravitational
constant, as predicted by some theories of gravity and current
cosmological models, leading to {\it gravitochemical heating}. If
only slow weak interactions are allowed in the neutron star
(modified Urca reactions, with or without Cooper pairing),
rotochemical heating can account for the observed ultraviolet
emission from the closest millisecond pulsar, PSR J0437-4715, which
also provides a constraint on $|dG/dt|$ of the same order as the
best available in the literature. 
\end{abstract}

\section{Introduction}
\label{sec:intro} Neutron star matter is composed of degenerate
fermions of various kinds: neutrons ($n$), protons ($p$), electrons
($e$), probably muons ($\mu$) and possibly other, more exotic
particles. (We refer to electrons and muons collectively as leptons,
$l$.) Neutrons are stabilized by the presence of other, stable
fermions that block (through the Pauli exclusion principle) most of
the final states of the beta decay reaction $n\to p+l+\bar\nu$.
The large chemical potentials $\mu_i$ ($\approx$ Fermi energies)
for all particle species $i$ also make inverse beta decays,
$p+l\to n+\nu$, possible. The neutrinos ($\nu$) and antineutrinos
($\bar\nu$) leave the star without further interactions,
contributing to its cooling (e.~g., \citealt{shapiro,yakovlev04}).
The two reactions mentioned tend to drive the matter into a
chemical equilibrium state, defined by
$\eta_{npl}\equiv\mu_n-\mu_p-\mu_l=0$.

If a matter element is in some way driven away from chemical
equilibrium ($\eta_{npl}\neq 0$), free energy is stored, which is
released by an excess rate of one reaction over the other. This
energy is partly lost to neutrinos and antineutrinos (undetectable
at present), and partly used to heat the matter. The heat is
eventually lost as thermal (ultraviolet) photons emitted from the
stellar surface.

The chemical imbalance can be caused by a change in the density of
the stellar matter. This can in turn be produced in different ways.
The first to be considered (by \citealt{finzi65,finzi_wolf}) was
stellar pulsation; however, so far no clear evidence for this
process has been seen. Gravitational collapse
\citep{haensel92,gourgoulhon93} and mass accretion are also possible
mechanisms, but in these contexts the non-equilibrium heating is
probably overwhelmed by the energy released through other channels.

Here we review our work on neutron star heating through beta
processes in two other contexts, which we consider to be the most
promising in revealing information about the physics of dense matter
and gravitation. One is {\it rotochemical heating}
(\citealt{reisenegger95,reisenegger97,fernandez05}, hereafter FR05;
\citealt{reisenegger06}), in which the precisely measurable decrease
of the stellar rotation rate, through the related reduction of the
centrifugal force, makes the star contract progressively, keeping it
away from chemical equilibrium. The more speculative {\it
gravitochemical heating} is based on the hypothesis that the
gravitational ``constant'' may in fact vary in time, causing a
similar contraction or expansion of the neutron star
(\citealt{jofre06}, hereafter JRF06). We refer to our published
papers for a detailed discussion of our methods and formalism (see
also \citealt{flores06}). Here we restrict ourselves to a general,
unified discussion of these two processes and their main
implications.

\section{Time evolution}
\label{sec:evolution}

The formalism for calculating the evolution of the temperature and
chemical imbalances for the case of rotochemical heating is
described in \S~2 of FR05. Here, we just outline the fundamental
equations and the modifications required in order to treat the
gravitochemical case as well. The evolution of the internal
temperature, $T$, taken to be uniform inside the star, is given by
the thermal balance equation,
\begin{equation}
\label{eq:dotT} \dot T=\frac{1}{C(T)}
\left[L_H(\eta_{npl},T)-L_{\nu}(\eta_{npl},T)-L_{\gamma}(T) \right],
\end{equation}
where $C$ is the total heat capacity of the star, $L_H$ is the total
power released by the heating mechanism, $L_{\nu}$ the total power
emitted as neutrinos, and $L_{\gamma}$ the power released as thermal
photons. Here and in what follows (including all figures), all
temperatures, chemical imbalances, stellar radii, and luminosities
are ``redshifted'' to the reference frame of a distant observer at
rest with respect to the star.

The evolution of the chemical imbalances is given by
\begin{eqnarray}
\label{eq:doteta} \dot{\eta}_{npl} & = & -
\left[A_{D,l}(\eta_{npe},T) +
A_{M,l}(\eta_{npe},T)\right]\nonumber \\
 & & - \left[B_{D,l}(\eta_{np\mu},T)+ B_{M,l}(\eta_{np\mu},T)\right]\nonumber \\
 & & - R_{npl}\Omega\dot\Omega + C_{npl} \dot G.
\end{eqnarray}
The functions $A$ and $B$ quantify the effect of reactions towards
restoring chemical equilibrium, and thus have the same sign of
$\eta_{npl}$ (FR05). The subscripts $D$ and $M$ refer to direct Urca
reactions,
\begin{eqnarray}
\label{eq:urca}
n & \longrightarrow & p + l + \overline{\nu}, \nonumber \\
p + l & \longrightarrow & n + \nu,
\end{eqnarray}
which are possibly forbidden by momentum conservation, and modified
Urca,
\begin{eqnarray}
\label{eq:murca}
n + N & \longrightarrow & p + N + l + \overline{\nu}, \nonumber \\
p + l + N  &  \longrightarrow & n + N + \nu,
\end{eqnarray}
where an aditional nucleon $N$ must participate in order to conserve
momentum (e.~g., \citealt{shapiro,yakovlev04}). The scalars
$R_{npl}$ and $C_{npl}$ quantify the departure from equilibrium due
to the changes in the centrifugal force ($\propto \Omega\dot\Omega$)
and the gravitational constant ($\dot G$), being positive and
depending on the stellar model and equation of state (FR05;
\citealt{reisenegger06}; JRF06).

\begin{figure}[!t]
  \includegraphics[width=\columnwidth]{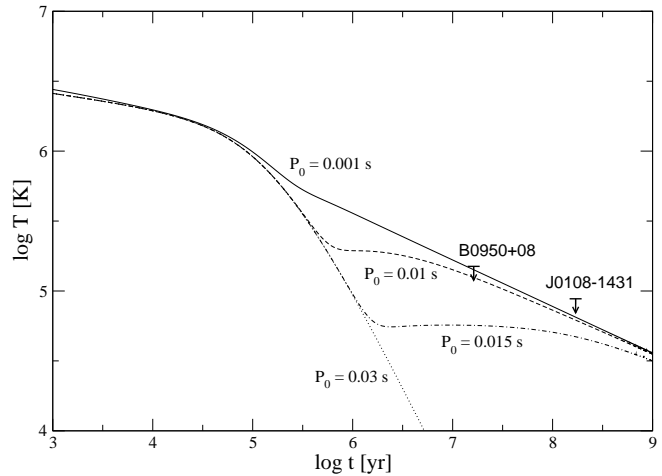} \caption{Predicted time-evolution of the surface
temperature, $T_s$, of a neutron star with rotochemical heating. All
curves correspond to stars with mass $M=1.4~M_\odot$, equation of
state A18$+\delta\upsilon+$UIX$^*$ \citep{apr98}, which allows only
modified Urca processes to occur, and magnetic dipole spin-down with
$B_{\mathrm{dipole}}=2.5\times 10^{11}~\mathrm{G}$. Each curve is
labeled by the assumed initial rotation period. The upper limits
correspond to observational constraints for pulsars B0950+08
\citep{zavlin04} and J0108$-$1431 (\citealt{mignani03}, as
interpreted by \citealt{kargaltsev04}), both of which have magnetic
dipole field strengths very close to the assumed value.}
\label{fig:evol_xi1}
\end{figure}

Figure~\ref{fig:evol_xi1} shows the solution of the coupled
differential equations \ref{eq:dotT} and \ref{eq:doteta} for the
evolution of a classical pulsar with a moderate magnetic field and
different assumed initial rotation periods under pure rotochemical
heating ($\Omega\dot\Omega<0$, $\dot G=0$). It can be seen that, for
very fast initial rotation, the pulsar can be kept warm beyond the
standard cooling time of $\sim 10^7~{\rm yr}$, at a level that is
close to current observational constraints.

\begin{figure}[!t]
  \includegraphics[width=\columnwidth]{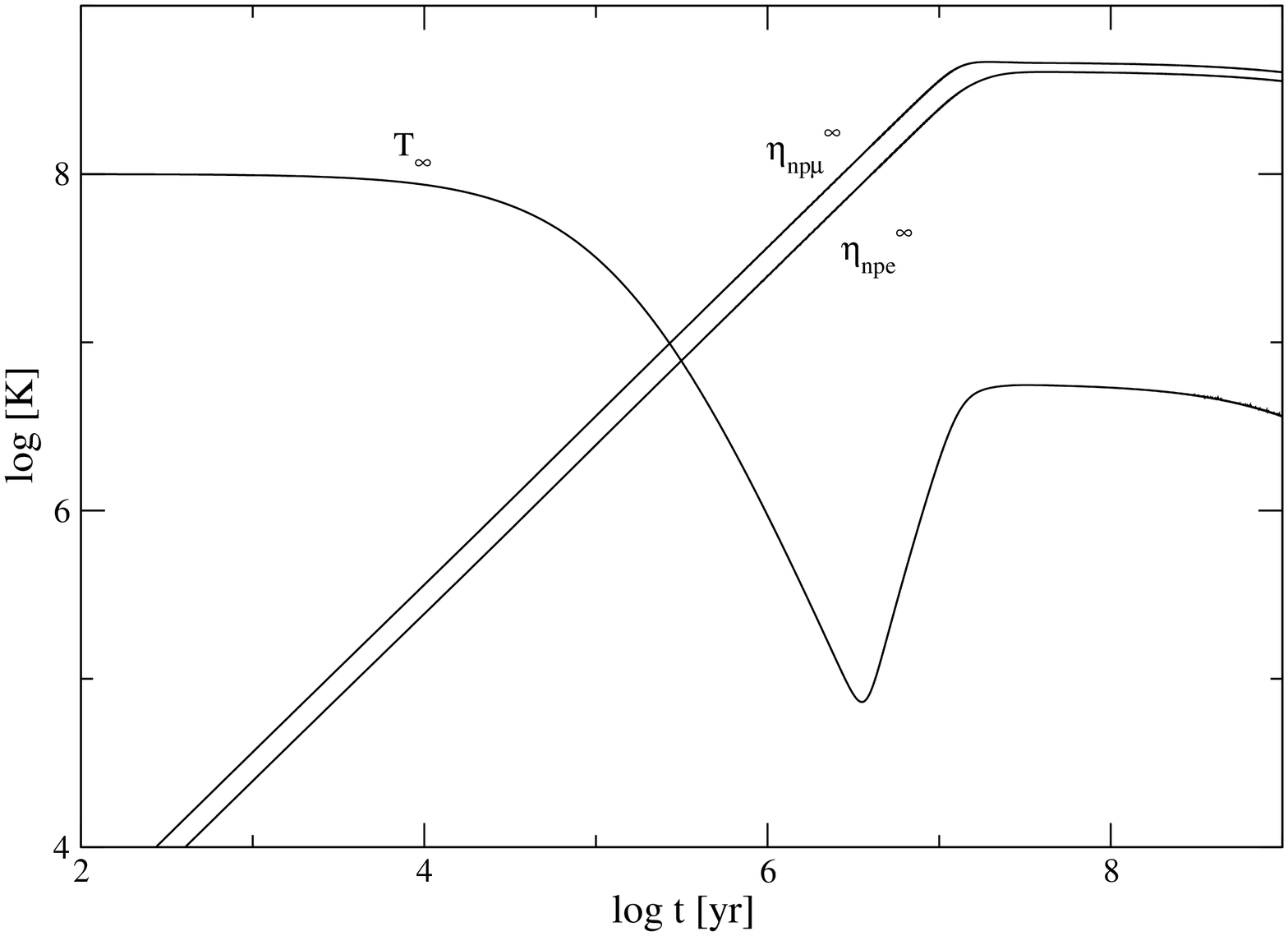} \caption{(taken from FR05)
Evolution of the internal temperature and chemical imbalances under
the rotochemical heating effect for a 1.4$M_\odot$ star calculated
with the A18 + $\delta \upsilon$ + UIX* equation of state
\citep{apr98}, with initial temperature $T = 10^8$ K, null initial
chemical imbalances, and magnetic dipole spin-down with field
strength $B=10^8$ G and initial period $P_0 = 1$ ms.}
\label{fig:evol_xi}
\end{figure}

The case of a ``millisecond pulsar'' (a neutron star with fast
rotation and a weak magnetic dipole field) is illustrated in
Figure~\ref{fig:evol_xi}. It first cools down from its high birth
temperature, while the chemical potential imbalances $\eta_{npl}$
slowly increase due to the decreasing rotation rate, until
rotochemical heating increases the temperature again, and the
reactions stop the rise of the chemical potential imbalances.

\section{Stationary state} \label{sec:stationary} 
\begin{figure}
  \includegraphics[width=\columnwidth]{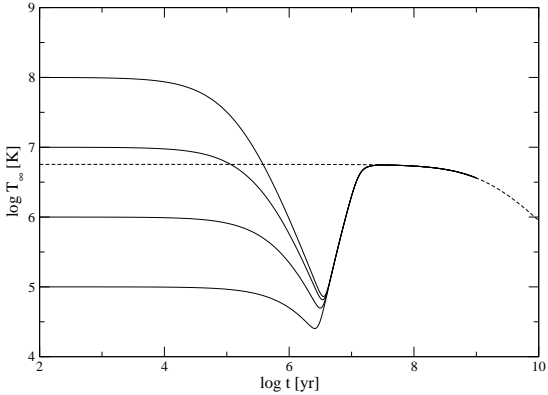} \caption{(from FR05) Evolution of the internal
temperature under rotochemical heating for different initial
temperatures. We set $\eta_{npe} = \eta_{np\mu}=0$. The short-dashed
line is the quasi-equilibrium solution, obtained by solving $\dot{T}
=0$ and $\dot{\eta}_{npe} = \dot{\eta}_{np\mu} =0$. The stellar
model and spin-down parameters are the same as in
Figure~\ref{fig:evol_xi}. \label{fig:evolTvarios}}
\end{figure}

\begin{figure}
  \includegraphics[width=\columnwidth]{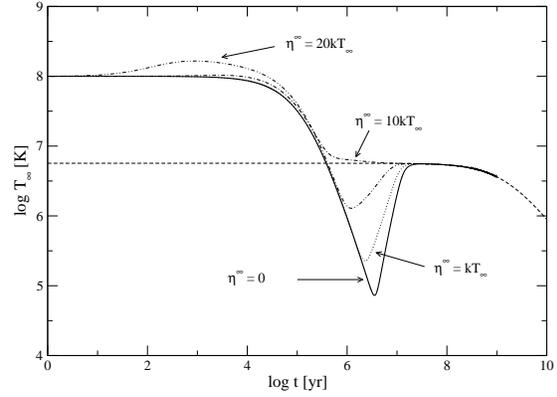} \caption{(from FR05) Evolution of the internal
temperature under rotochemical heating for different initial
chemical imbalances $\eta_{npe} = \eta_{np\mu} \equiv \eta$ and the
same initial temperature $T = 10^8$ K at $t=0$. The line styles, the
stellar model, and the spin-down parameters are the same as in
Figure~\ref{fig:evolTvarios}. \label{fig:evolmuvarios}}
\end{figure}

If the relevant
forcing ($\Omega\dot\Omega$ or $\dot G$) changes slowly with time,
the star eventually arrives at a stationary state, where the rate at
which the equilibrium concentrations are modified by this forcing is
the same as that at which the reactions drive the system toward the
new equilibrium configuration, with heating and cooling balancing
each other \citep{reisenegger95}. The evolution to this state for
pure rotochemical heating is illustrated in
Figure~\ref{fig:evol_xi}. Figures~\ref{fig:evolTvarios} and
\ref{fig:evolmuvarios} show that the state reached is independent on
the assumed initial conditions.

The properties of this stationary state can be obtained by the
simultaneous solution of equations (\ref{eq:dotT}) and
(\ref{eq:doteta}) with $\dot T = \dot \eta_{npl} = 0$. The existence
of the stationary state makes it unnecessary to model the full
evolution of the temperature and chemical imbalances of the star in
order to calculate the final temperature, since the stationary state
is independent of the initial conditions (see FR05 for a detailed
analysis of the rotochemical heating case). For given values of
$\Omega\dot\Omega$ and $\dot G$, it is thus possible to calculate
the temperature of an old pulsar that has reached the stationary
state, without knowing its exact age.

When only modified Urca reactions operate, it is possible to solve
analytically for the stationary values of the photon luminosity
$L_{\gamma}^{st}$ and chemical imbalances $\eta_{npl}^{st}$, as a
function of stellar model and current value of $\Omega\dot\Omega$
and $\dot G$. The reason for this is that the longer equilibration
timescale given by the slower modified Urca reactions yields
stationary chemical imbalances satisfying $\eta_{npl}\gg kT$. In
this limit, the term $L_H-L_{\nu}$ in the thermal balance equation
can be written as $K_{Le}\eta_{npe}^8 + K_{L\mu} \eta_{np\mu}^8$,
where $K_{L,l}$ are positive constants that depend only on stellar
mass and equation of state (FR05; JRF06). For typical equations of
state, the photon luminosity in the stationary state is
\begin{equation}
L_{\gamma}^{st} \simeq 10^{30-31} \left|{\dot P_{-20}\over P_{\rm
ms}^3}+{\dot G/G\over 3\times 10^{-11} \textrm{\,\
yr}^{-1}}\right|^{8/7} \textrm{erg s}^{-1},
\end{equation}
where $P_{\rm ms}$ is the rotation period in milliseconds, and $\dot
P_{-20}$ is its time derivative in units of $10^{-20}$
(dimensionless), and the effective surface temperature of the star
in the stationary state is
\begin{equation}
T_{s}^{st} \simeq (2-3) \times 10^5 \left|{\dot P_{-20}\over P_{\rm
ms}^3}+{\dot G/G\over 3\times 10^{-11} \textrm{\,\
yr}^{-1}}\right|^{2/7} \textrm{K}.
\end{equation}
Finally, the timescale for the system to reach the stationary state
is
\begin{equation} \label{eq:tau_eq}
\tau_{st} \simeq 2 \times 10^7 \left|{\dot P_{-20}\over P_{\rm
ms}^3}+{\dot G/G\over 3\times 10^{-11} \textrm{\,\
yr}^{-1}}\right|^{-6/7} \textrm{yr}.
\end{equation}

\section{Comparison to observations}
\label{sec:comparison} In order to verify this model and constrain
the value of $|\dot G/G|$, we need a neutron star that (1) has a
measured surface temperature (or at least a good enough upper limit
on the latter), and (2) is confidently known to be older than the
timescale to reach the stationary state. So far, the only object
satisfying both conditions is the millisecond pulsar closest to the
Solar System, PSR~J0437-4715 (hereafter J0437), whose surface
temperature was inferred from an HST-STIS ultraviolet observation by
\citet{kargaltsev04}. Its spin-down age, $\tau_\mathrm{sd} \simeq 5
\times 10^{9}$~yr (e.g., \citealt{vbb01}), and the cooling age of
its white dwarf companion, $\tau_\mathrm{WD}\simeq (2.5-5.3)\times
10^{9}$~yr~\citep{hansen98}, are much longer than the time required
to reach the steady state for both rotochemical and gravitochemical
heating, in the latter case under the condition that $|\dot G/G|\geq
10^{-13}$ yr$^{-1}$.

\begin{figure}[!t]
  \includegraphics[width=\columnwidth]{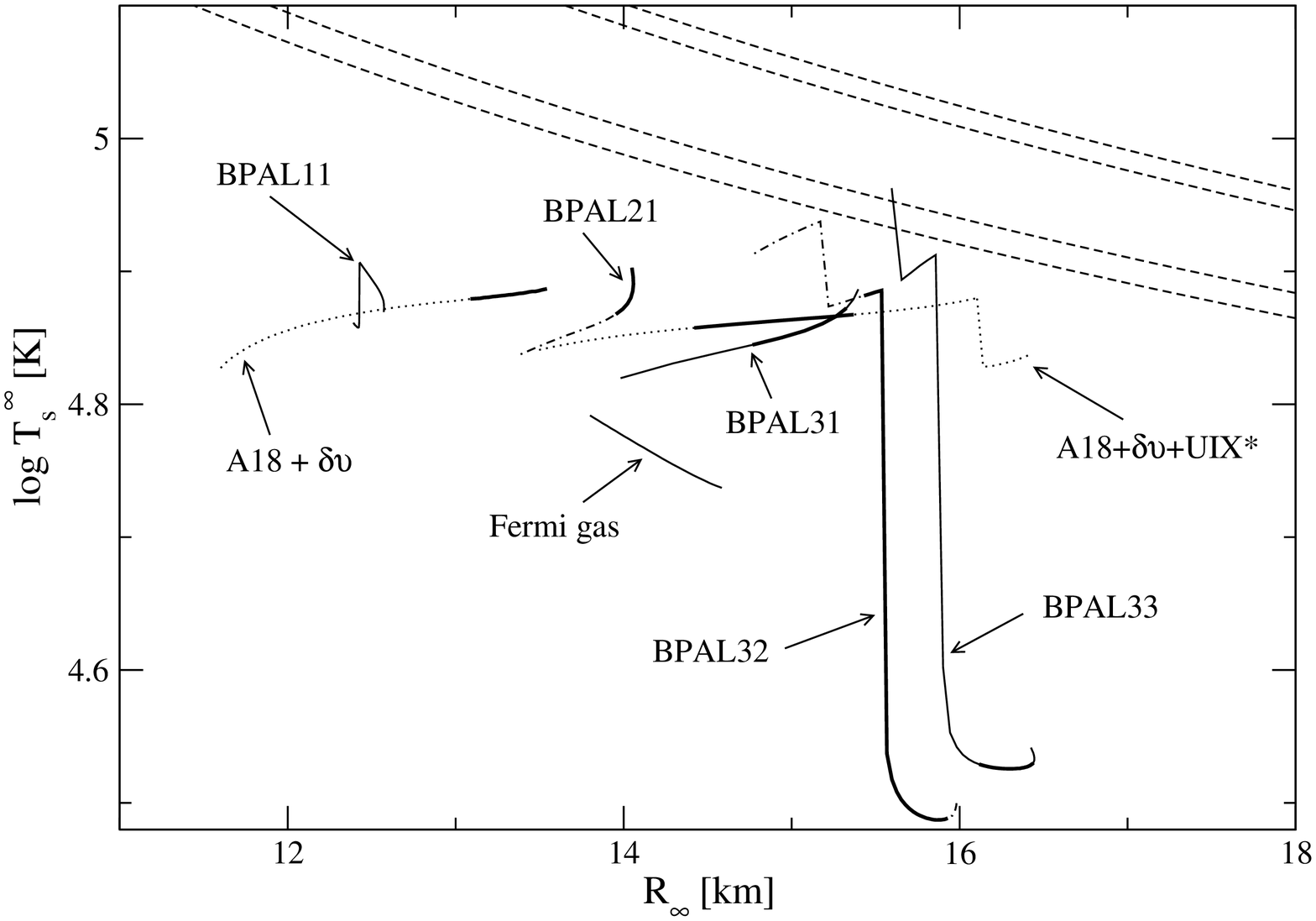} \caption{(taken from FR05)
Surface temperature due to rotochemical heating in the stationary
state as function of stellar radius for different equations of
state, shown as solid lines (APR from \citealt{apr98} and BPAL from
\citealt{pal88}), for the spin parameters of PSR J0437-4715. Dashed
lines are 68\% and 90\% confidence contours of the blackbody fit to
the emission from this pulsar \citep{kargaltsev04}. Bold lines
indicate, for each equation of state, the mass range allowed by the
constraint of \citet{vbb01}, $M_{PSR} = 1.58\pm 0.18M_\odot$. BPAL32
and BPAL33 allow direct Urca reactions in the observed mass range of
PSR~J0437.} \label{fig:Ts_Rinf}
\end{figure}

Consequently, we consider stellar models constructed from different
equations of state, with masses satisfying the constraint obtained
for J0437 by \citet{vbb01}, $M_\mathrm{psr} = 1.58 \pm 0.18
M_{\odot}$, and calculate the stationary temperature for each.
Figure~\ref{fig:Ts_Rinf} compares the predictions for the case of
pure rotochemical heating with the measured spin-down parameters of
J0437 (for various equations of state and neutron star masses) to
the temperature inferred from the observation of
\citet{kargaltsev04}.

\begin{figure}
\includegraphics[scale=0.525]{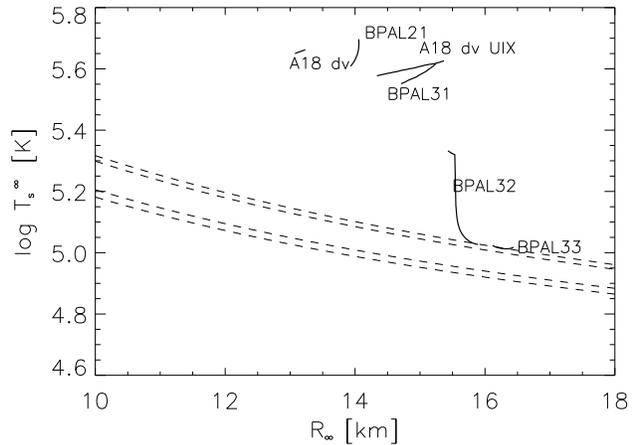}
\caption{(from JRF06) Surface temperature due to gravitochemical
heating in the stationary state as function of stellar radius for
different equations of state. The value of $|\dot{G}/G| = 2 \times
10^{-10}$~yr$^{-1}$ is chosen so that all stationary temperature
curves lie above the observational constraints. Otherwise, the
meanings of lines and symbols are as in Figure~\ref{fig:Ts_Rinf}.
\label{fig:J0437_1}}
\end{figure}

In Figure~\ref{fig:J0437_1}, we compare the same observational
constraints on the temperature of this pulsar to the theoretical
predictions for pure gravitochemical heating, assuming $|\dot{G}/G|
= 2\times 10^{-10}$~yr$^{-1}$. As can be seen, this value is such
that the stationary temperatures of all stellar models lie just
above the 90 \% confidence contour, and therefore represents a
rather safe and general upper limit.

\begin{figure}
\includegraphics[scale=0.525]{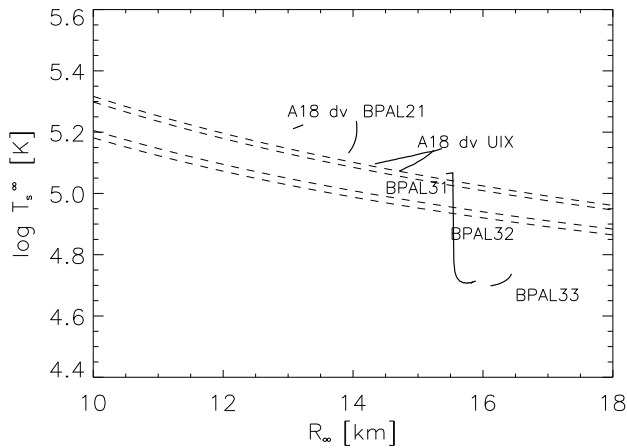}
\caption{(from JRF06) Same as Figure~\ref{fig:J0437_1}, but now the
value of $|\dot{G}/G| = 4 \times 10^{-12}$ yr$^{-1}$ is chosen such
that only the stationary temperature curves with modified Urca
reactions are above the observational constraints.
\label{fig:J0437_2}}
\end{figure}

When the stellar mass becomes large enough for the central
pressure to cross the threshold for direct Urca reactions, $T_s$
drops abruptly, due to the faster relaxation towards chemical
equilibrium. This occurs in two steps, as electron and muon direct
Urca processes have different threshold densities (see, e.g.,
FR05). Conventional neutron star cooling models reproduce observed
temperatures better when only modified Urca reactions are
considered (e.~g., \citealt{yakovlev04,page06}). Restricting our
sample to the equations of state that allow only modified Urca
reactions in the mass range considered here, namely $A18 + \delta
v$, $A18 + \delta v +$ UIX, BPAL21, and BPAL31, we obtain a more
restrictive upper limit on $|\dot{G}|$, as shown in
Figure~\ref{fig:J0437_2}, yielding $| \dot G / G | < 4 \times
10^{-12}$ yr$^{-1}$.

\section{Discussion and Conclusions}
\label{sec:discussion}

\subsection{Rotochemical heating}

Using the equations of state that allow only modified Urca reactions
within the allowed mass for PSR J0437-4715, rotochemical heating
predicts an effective temperature in the narrow range $T_{s,eq} =
(6.9 - 7.9)\times 10^4$ K, about 20\% lower than the blackbody fit
of \citet{kargaltsev04}. There are three possible reasons why the
prediction does not quite match the observation:
\begin{enumerate}
\item  We are not taking superfluidity into account. This would reduce Urca reaction rates,
lengthening the equilibration timescale and raising the
stationary-state temperature \citep{reisenegger97}.

\item We are neglecting other heating mechanisms (some of them
directly related to superfluidity), which could further raise the
temperature at any stage in the thermal evolution. Nonetheless, in
millisecond pulsars, all proposed mechanisms appear to be less
important than rotochemical heating \citep{schaab99,kargaltsev04}.

\item The thermal spectrum could deviate from a blackbody, as for
the isolated neutron star RX J1856-3754, which has a
well-determined blackbody X-ray spectrum that underpredicts the
optical flux \citep{walter}, indicating a more complex spectral
shape of its thermal emission.

\end{enumerate}

\citet{kargaltsev04} stress that PSR J0437-4715 has a higher
surface temperature than the upper limit for the younger,
``classical'' pulsar J0108-1431, $T_s<8.8\times 10^4$ K, inferred
from the optical non-detection by \citet{mignani03} and shown in
our Figure~\ref{fig:evol_xi1}. In the rotochemical heating model,
these two pulsars are in very different regimes: J0437 is in the
stationary state in which its temperature can be predicted from
its spin-down parameters, whereas J0108 has a 680 times smaller
spin-down power ($\propto\Omega\dot\Omega$), and will therefore
not reach a detectable stationary state. Its equilibration
timescale, according to equation (\ref{eq:tau_eq}), is $2\times
10^{11}$ yr, longer than the age of the Universe and certainly
much longer than the spin-down age of the pulsar. Thus, its heat
content (if any) is due to its earlier, faster rotation, which may
have built up a significant chemical imbalance that is currently
being decreased by ongoing reactions in its interior (see
Fig.~\ref{fig:evol_xi1}). Depending on its initial rotation
period, its surface temperature may be substantially smaller than
both J0437's observed temperature and its own current upper limit.

\subsection{Gravitochemical heating}

\begin{table*}[t]
\caption{Previous upper bounds on $|\dot G / G|$.}
\centering
\label{tab:1}       
\begin{tabular}{ c  c  c  c } \hline
Method  & $|\dot{G}/G|_{\rm max}$ [$10^{-12}~\mathrm{yr}^{-1}$]& Time scale [yr] & Reference \\
\hline \hline
Big Bang Nucleosynthesis  & $0.4$    & $1.4\times 10^{10}$ & \citet{copi04} \\
Microwave background &   $0.7$    &  $1.4\times 10^{10}$ & \citet{nagata04}   \\
Globular cluster isochrones &   $35$    & $10^{10}$ &  \citet{innocenti96} \\
Binary neutron star masses  &   $2.6$ & $10^{10}$ &  \citet{thorsett96} \\
Helioseismology &   $1.6$    & $4\times 10^9$ & \citet{guenther98} \\
Paleontology   &   $20$  & $4\times 10^9$   &   \citet{eichendorf77} \\
Lunar laser ranging   &   $1.3$    & 24 & \citet{williams04}   \\
Binary pulsar orbits   &   $9$    & $8$ &   \citet{kaspi94}  \\
White dwarf oscillations   &   $250$ & $25$ &  \citet{benvenuto04}    \\
\hline
\end{tabular}
\end{table*}

Table 1 lists some of the many experiments performed so far to
test the constancy of $G$ (see \citealt{uzan} and \citealt{will}
for recent reviews). The second column contains the upper limits
on its time variation, most usefully expressed as $| \dot G / G
|$, and the third is a rough time scale over which each
experiment is averaging this variation. Based on the latter, the
experiments can be separated into three classes. The first two
experiments on the list measure the variation of $G$ from the
early Universe to the present time,
and the constraint on the present-day value of $|\dot G/G|$ is
based on assuming a time dependence $G(t)\propto t^{-\alpha}$,
where $t$ is the time since the Big Bang, and $\alpha$ is a
constant constrained by these experiments. The next four are
sensitive to variations over long timescales, $10^{9-10}$~yr, but
without reaching into the very early Universe. The last four
experiments measure the change of $G$ directly over short,
``human'' timescales of years or few decades. Even though results
from the first category are nominally the most restrictive on a
long-term variation of $G$, they depend crucially on the assumed
form of the variation of $G$ near the Big Bang. Thus, it is still
useful to consider measurements of the second and third
categories, which could directly detect variations of $G$ in more
recent times.

The new method advocated here, namely gravitochemical heating of
neutron stars, falls closest to the second category, as its
timescale is much longer than human, but does not reach into the
early Universe. However, it probes somewhat shorter timescales than
the other methods in this category. In the most general case, when
direct Urca reactions are allowed to operate, we obtain an upper
limit $|\dot G / G| < 2 \times 10^{-10}$ yr$^{-1}$. Restricting the
sample of equations of state to those that allow only modified Urca
reactions, we obtain a much more restrictive upper limit, $|\dot G /
G| < 4 \times 10^{-12}$ yr$^{-1}$ on a time scale $\sim 10^8~{\rm
yr}$ (the time for the neutron star to reach its quasi-stationary
state), competitive with constraints obtained from the other methods
probing similar timescales. However, since the composition of matter
above nuclear densities is uncertain and millisecond pulsars are
generally expected to be more massive than classical pulsars, we
cannot rule out the result for the direct Urca regime.

Further progress in our knowledge of neutron star matter will
allow this method to become more effective at constraining
variations in $G$. The method can also be improved with an
increased sample of objects with measured thermal emission or good
upper limits.

\begin{acknowledgement}
We thank G.~Pavlov and O.~Kargaltsev for letting us know about their
work in advance of publication and for kindly providing their data
in electronic form. The authors are also grateful to C.~Dib,
O.~Espinosa, S.~Flores-Tuli\'an, M.~Gusakov, E.~Kantor, R.~Mignani,
D.~Page, M.~Ta\-ghi\-za\-deh, and M.~van Kerkwijk for discussions
that benefited the present paper. This work made use of NASA's
Astrophysics Data System Service, and received financial support
from FON\-DE\-CYT through regular grants 1020840 and 1060644.
\end{acknowledgement}

\end{document}